\documentclass{JHEP3}

\usepackage{amsmath}
\usepackage{amssymb}
\usepackage{mathrsfs}

\def\eq#1 { \begin{equation} #1 \end{equation} }
\def\eqn#1{ \begin{eqnarray} #1 \end{eqnarray} }
\def\nn { \nonumber }

\def \cH {\mathcal{H}}
\def \cHa { \mathcal{H}_{\rm aux} }
\def \cHp { \mathcal{H}_{\rm phys} }
\def \cHl{\cH_{\rm L}}
\def \cHr{\cH_{\rm R}}

\def\aux{[\!]}                                  
\def\dket#1{\aux #1 \rangle}                    
\def\dbra#1{\langle #1 \aux}                    
\def\ket#1{\vert #1 \rangle}                    
\def\brak#1#2{\langle #1 \vert #2 \rangle}      
\def\sket#1{\aux #1 \rangle_{\rm seed}}         
\def\sbra#1{{}_{\rm seed}\langle #1 \aux}       
\def\sbrak#1#2{ {}_{\rm seed}\langle #1 \aux #2 \rangle_{\rm seed} }

\def \half {\frac{1}{2}\,}

\def \xp {x^+}
\def \xm {x^-}
\def \t  {\tau}
\def \th {\theta}
\def \ph {\phi}
\def \a  {\alpha}
\def \at {\widetilde{\alpha}}
\def \d  {\partial}
\def \Lt {\widetilde{L}}
\def \h  {\widetilde{h}}
\def \nt {\widetilde{n}}

\def \Int {\mathbb{Z}}
\def \Reals {\mathbb{R}}
\def \rpi {\sqrt{\pi}}
\def \Proj { \mathcal{P}_0 }
\def \Lie {\pounds}

\def \cNs {\mathcal{N}_{\rm seed}}
\def \cNa {\mathcal{N}_{\rm aux}}
\def \dim {{\rm dim}}
\def \Ss  {S_{\rm seed}}
\def \Sa  {S_{\rm aux}}

\def \vac {\dket{0}}
\def \hvac {\dket{0;h,\h}}
\def \ovac {\dket{0;0,0}}
\def \hnket {\dket{n,\nt;h,\h}}

\def \com#1#2{ \left[ #1, #2 \right] }

\title{Group Averaging of massless scalar fields in $1+1$ de Sitter}

\author{Donald Marolf and Ian A Morrison \\ 
  Department of Physics, University of California at Santa
  Barbara, Santa Barbara, CA 93106, USA \\
  {\it E-mail:} \email{marolf@physics.ucsb.edu}, 
  \email{ian\_morrison@physics.ucsb.edu}
}

\abstract{
  Perturbative gravity in global de Sitter space is subject
  to so-called linearization stability constraints: If they are to
  couple consistently to the gravitational field, quantum states
  must be invariant under the de Sitter isometries.
  While standard Fock spaces contain no de Sitter-invariant states 
  apart from (possibly) the vacuum, a full Hilbert space of de Sitter-invariant
  quantum states can be constructed via group averaging techiniques.
  We re-examine the simple toy model of de Sitter group averaging given by the
  free 1+1 scalar field, expanding on an earlier
  analysis by Higuchi.  Our purpose is twofold:  to include the scalar
  zero-mode, and to explicitly count the number of de Sitter-invariant states
  as a function of an appropriately defined energy.
}
\preprint{}

\begin{document}


\section{Introduction}

Understanding quantum gravity in de Sitter space remains an
important problem.  A major motivation is the relevance of de Sitter
to cosmology: measurements of the CMB \cite{WMAP} are consistent
with a period of inflation in which the universe underwent a de
Sitter-like phase of rapid expansion, and observations of type Ia
supernovae suggest \cite{SN} that our universe may have a small
positive cosmological constant and may approach de Sitter space in
the far future.  Thus any theory of quantum gravity should include a
description of de Sitter space, at least in some approximate form.
Unfortunately, the study of de Sitter quantum gravity has been
fraught with conceptual difficulties (see, e.g.,
\cite{Rees97,Banks00,Witten01}). In this paper we examine one
particular hurdle that arises in perturbative gravity about a de
Sitter background.

To summarize this hurdle, recall that field theories on spacetimes
with Killing symmetries have conserved charges. We wish to
regard such a theory (together with linearized gravitational waves)
as the zero-order perturbative approximation to a theory of matter
plus gravity.  This context is particularly interesting when the
background also has compact Cauchy surfaces.  Then the gravitational
equivalent of Gauss' law implies that the above charges must vanish
in order for a solution to this zero-order theory to consistently
couple  to dynamical gravity
\cite{BrillDeser73,Moncrief75,Moncrief76,Moncrief78,Arms77,
  FisherMarsden79,ArmsEtal82}.
Since these constraints are not encoded in the linearized field
equations, they are known as linearization-stability constraints.

In de Sitter space, the linearization-stability constraints require
linearized quantum states to be invariant under the de Sitter group
$SO_0(D,1)$ where $D$ is the spacetime dimension
\cite{Higuchi91a,Higuchi91b,LosicUnruh06}. Because the de Sitter
group is non-compact, the standard Fock space contains no de
Sitter-invariant states except for a possible vacuum
\cite{Higuchi91b}. This meager set of states is clearly insufficient
to reproduce the rich physics of the corresponding classical theory.
Fortunately, however, one may use the standard Fock space (which we
call the `auxiliary' Hilbert space $\cHa$) to build a new `physical'
Hilbert space $\cHp$ of de Sitter-invariant states via group
averaging \cite{Higuchi91b}\footnote{See \cite{KL,Marolf95} for
independent introductions of similar techniques in related
contexts.}. This technique considers linear superpositions of
auxiliary states $\dket{\psi}$ of the form 
\eq{
  \label{eq:GAState}
  \ket{\Psi} := \int_{g\in G} dg\, U(g) \dket{\psi},
}
where $G$ is the de Sitter group, $dg$ is the unique (unimodular,
left- and right-invariant) Haar measure of $G$, and $U(g)$ gives the
unitary representation of $G$ on $\cHa$. Such superpositions are
formally invariant under the de Sitter group.  For compact groups
the analogue of (\ref{eq:GAState}) converges and gives the
projection of $\dket{\psi}$ onto the trivial representation.
However, since our $G$ is non-compact, the state (\ref{eq:GAState})
is not normalizable in $\cHa$. Nevertheless it can be understood
(see e.g. \cite{ALMMT}) as a ``generalized state'' in a sense
similar to that used for non-normalizable eigenstates of operators
with continuous spectrum (e.g., plane waves in infinite space).

More concretely, one defines a new inner product on the
group-averaged states (\ref{eq:GAState}):
 \eq{
 \label{eq:GAInnerProduct}
  \brak{\Psi_1}{\Psi_2} := \dbra{\psi_1} \! \cdot \! \ket{\Psi_2}
  = \int_{g\in G} dg\, \dbra{\psi_1} \ U(g) \ \dket{\psi_2} .
} 
The linear superposition (\ref{eq:GAState}) is meaningful when
this ``group-averaging inner product'' converges. 

When the sense of the convergence is sufficiently strong, a theorem
of \cite{GiuliniMarolf99} states that the group-averaging inner product
is the unique inner product consistent with the $\star$-algebra of 
bounded gauge-invariant observables in $\cHa$. More formal discussions 
of group averaging can be found in \cite{Marolf00,GiuliniMarolf98}.
Other studies of de Sitter group averaging include
\cite{GiddingsMarolf07,MarolfMorrison08}.

The purpose of this paper is to analyze group averaging for a
massless scalar field in $1+1$ de Sitter, completing the analysis
begun by Higuchi \cite{Higuchi91b}.  Higuchi was primarily concerned
with de Sitter group averaging for 3+1 gravitons
 and used the $1+1$ massless scalar as a
 toy model.  For simplicity, he omitted the scalar zero mode
 (which has no analogue for gravitons).
However, the physical massless scalar has a zero mode that should be
included in a more complete analysis.  We do so below. We also
compute the number of de Sitter-invariant states as a function of
energy flux through the de Sitter neck.  For energies much greater
than the de Sitter scale, a straightforward calculation shows that
this entropy agrees with that of the naive auxiliary Hilbert space.
This provides an explicit check of the argument presented in
\cite{GiddingsMarolf07} that such a result should hold for generic
field theories.

This paper is organized as follows. Section \ref{Sec:merge} briefly
reviews the quantization of the massless scalar in de Sitter.
Section \ref{sec:groupAveraging} then follows Higuchi in using group
averaging to construct an orthonormal basis of physical states from
special auxiliary ``seed states.'' The physical entropy is computed
in \ref{sec:entropy} and section \ref{sec:discussion} presents some
final discussion.

\section{Free scalar field in $1+1$ de Sitter}
\label{Sec:merge}

We begin with a brief overview of massless scalar fields in $1+1$ de
Sitter \cite{Allen87,Mottola84,Polarski90}. It is useful to adopt
conventions of conformal field theory
\cite{DiFrancesco97,Ginsparg88}, and to write the 1+1 de Sitter
metric in the form \eq{ \label{eq:metric}
  ds^2 = \frac{\ell^2}{\cos^2\tau}(- d\tau^2 + d\th^2 ) ,
} which is just a conformal factor times the metric on the cylinder.
Here the conformal time $\tau$  has range $-\pi/2 < \tau <
\pi/2$, $\th$ periodic $\th \cong \th + 2\pi$, and $\ell$ is the de
Sitter length scale.  We adopt lightcone coordinates $ x^\pm = \tau
\pm \th$.

\subsection{Operators and States}
\label{sec:scalar}

The action of a free scalar field is
\eq{ \label{eq:action}
  S = - \half \int d^2x \sqrt{-g} g^{ab} \nabla_a \ph \nabla_b \ph
  = \int d^2x\, \d_+\ph \d_- \ph ,
} where $g_{ab}$ is the de Sitter metric and $\nabla_a$ the
covariant derivative associated with $g_{ab}$. In the second
equality we note that the conformal factor from (\ref{eq:metric})
cancels out of the action, making the theory conformally invariant.
The equation of motion for $\ph$ is thus $\d_+ \d_- \phi(x) = 0$,
and the solutions are familiar left- and right-moving modes \eq{
  \d_+ \ph(\xp) = \frac{1}{2 \rpi} \sum_m \a_m \exp\left[- i m \xp \right] ,
  \quad
  \d_- \ph(\xm) = \frac{1}{2 \rpi} \sum_m \at_m \exp\left[- i m \xm \right]
  .
} Upon integrating one finds 
\eqn{ \label{eq:modeExpansion1}
  \ph(x) 
  &=& \frac{\ph_0}{4\pi} + \a_0 \xp + \at_0 \xm
  + \frac{i}{2 \rpi} \sum_{m \neq 0} \left[
    \frac{\a_m}{m} e^{-i m \xp} + \frac{\at_m}{m} e^{-i m \xm}\right] \nn \\
  &=& \frac{\ph_0}{4\pi} + (\a_0 + \at_0) \t + (\a_0 - \at_0) \th 
  \nn \\ & &
  + \frac{i}{2 \rpi} \sum_{m \neq 0} \frac{1}{m} e^{-i m \t}
    \left[ \a_m e^{-i m \th} + \at_m e^{+ i m \th} \right] .
}
We identify the term linear in $\tau$ as the momentum
$ p \propto (\a_0 + \at_0) $. The fact that $\ph(x)$ must be
single-valued places further constraints on the mode expansion,
depending on the target space of $\ph(x)$. We consider two cases:
\begin{enumerate}
\item[i)]
  The target space of $\ph(x)$ is the real line. Single-valuedness
  of $\ph(x)$ requires $\ph(\tau, \th+2\pi) = \ph(\tau,\th)$;
  thus $\a_0 = \at_0$ and the term linear in $\th$ in
  (\ref{eq:modeExpansion1}) vanishes.
\item[ii)]
  The target space of $\ph(x)$ is the circle $S^1$ with radius $R$.
  Single-valuedness requires $\ph(\tau,\th+2\pi) = \ph(\tau,\th) + 2\pi R w$,
  where $w \in \Int$ is the winding number of the field. From
  (\ref{eq:modeExpansion1}) we see that $ R w$ is given by
  $R w = (\a_0 - \at_0)$. Furthermore, because $\ph(x)$ is periodic,
  $p$ is quantized: $p = k / R, \; k \in \Int$. Periodic scalars in
  de Sitter have previously been considered in, e.g., \cite{JoungEtal07}.
\end{enumerate}
For the remainder of this section we will keep $p$ explicit so that
our expressions apply to either case; later we will specialize to
case (ii) and write expressions in terms of $k$. Our mode expansion
is now
\eq{ \label{eq:modeExpansion2}
  \ph(x) = \frac{\ph_0}{4\pi} + 2 p \tau + R w \th
  + \frac{i}{2 \rpi} \sum_{m \neq 0} \frac{1}{m} e^{-i m \t}
  \left[ \a_m e^{- i m \th} + \at_m e^{i m \th} \right] .
}

We now quantize our scalar field using canonical techniques, the end
result of which is the auxiliary Hilbert space $\cHa$. The
quantities $\ph_0$, $p$, $w$, $\a_m$, and $\at_m$ are promoted to
operators. Imposing the canonical commutation relation $[ \ph(\tau,
\th_1), \ph(\tau, \th_2) ] = i \delta(\th_1 -\th_2)$ we find \eq{
  \com{\ph_0}{p} = i, \quad
  \com{\a_m}{\a_n} = \com{\at_m}{\at_n} = m \delta_{m,-n} ,
}
with all other commutators vanishing. In the usual fashion, $\a_m$ and
$\at_m$ are interpreted as left- and right- moving creation operators
($m < 0$) and annihilation operators ($m > 0$). It will be useful
to use the Virasoro generators $L_0,\,L_{\pm 1}$
\cite{DiFrancesco97,Ginsparg88}
\eq{
  L_m = \half \sum_{n=-\infty}^{\infty} :\a_{m-n} \a_n:,
}
which obey the algebra
\eq{
  \com{L_{\pm 1}}{L_0} = \pm L_{\pm 1}, \quad
  \com{L_1}{L_{-1}} = 2 L_0,
}
and likewise for $\Lt_0,\,\Lt_{\pm 1}$.

We can define a vacuum state $\vac$ as the state for which
\eq{ \label{eq:vacua}
  \a_m \vac = \at_m \vac = 0 \quad \forall\;  m > 0 .
} Such a vacuum state is not in general annihilated by $\a_0$ or
$\at_0$.  Instead, there is a two-parameter family of vacua
distinguished by their eigenvalues of $\a_0$ and $\at_0$, i.e. the
momentum and winding of each vacuum. It is equivalent to label
independent vacua by their eigenvalues $h$ and $\h$ of the Virasoro
generators $L_0$ and $\Lt_0$: \eq{
  h = \half \left( p + \frac{R w}{2} \right)^2 , \quad
  \h = \half \left( p - \frac{R w}{2} \right)^2 ;
} we therefore denote a vacuum by $\hvac$. We shall see shortly that
the only de Sitter-invariant vacuum is the $p=w=0$ vacuum $\ovac$.
Excited states are created by acting on a vacuum with creation
operators $\a_m$ ($\at_m$) for $m < 0$, and will be labeled using
the somewhat degenerate notation $\hnket$, where $n, \tilde n$ are
the eigenvalues of $L_0 -h, \Lt_0 - \tilde h$ and we refer to $N :=
n + \nt$ as the \emph{level} of a state.  Each creation operator
$\a_m$ ($\at_m$) increases the eigenvalue of $L_0$ ($\Lt_0$), and
thus the level, by $m$.

Let us also introduce the operator \eq{
  H = L_0 + \Lt_0 ,
} which generates translations in $\tau$; i.e., it is the
Hamiltonian for the conformally rescaled problem on the cylinder
$S^1\times\Reals$, up to a constant offset associated with the
Casimir energy. Since de Sitter space does not have a global
future-directed timelike Killing field, $H$ is not naturally thought
of as a de Sitter Hamiltonian. However, it does agree with the flux
of de Sitter stress-energy through the sphere at $\tau =0$ (again up
to a constant offset).  In this latter form, this operator was an
important ingredient in the analysis of \cite{GiddingsMarolf07}.  We
shall thus refer to $H$ as an ``energy.''  This operator acts on a
state $\hnket$ as \eqn{
  H \hnket
  &=& \left( h + \h + n + \nt \right) \hnket \nn \\
  &=& \left( p^2 + \frac{R^2 w^2}{4} + n + \nt \right)
  \hnket ,
  \label{eq:H}
} and so the energy of such a state is $E := h + n + \h + \nt$.

\subsection{The de Sitter group} \label{sec:symmetry}

Let us quickly review the symmetries of $1+1$ de Sitter spacetime.
This space has three independent Killing vector fields which we may
take to be \eq{ \label{eq:dSGenerators}
  \d_\th = \half (\d_+ - \d_- ), \quad
  \xi_1^a \d_a = \half ( \cos( \xp) \d_+ + \cos( \xm)  \d_- ) , \quad
  \xi_2^a \d_a = \half ( \sin( \xp) \d_+ + \cos( \xm)  \d_- ) .
}
 Such isometries can be understood by embedding 1+1 de Sitter
 in $2+1$
Minkowski space: there $\d_\th$ generates rotations preserving the
Cartesian coordinate $X^0$, while $\xi_1^a$ and $\xi_2^a$ generate
boosts along the Cartesian spatial directions. The Killing fields
act on $\cHa$ via operators $J$, $B_1$, and $B_2$ which satisfy the
$SO_0(2,1)$ algebra \eq{
  \com{B_1}{B_2} = i J, \quad
  \com{B_1}{J} = i B_2, \quad
  \com{B_2}{J} = -i B_1 .
} On the scalar field $\ph(x)$, their action is \eq{
\label{eq:generatorAction}
  \com{B_1}{\phi(x)} = i \Lie_{\xi_1} \phi(x) = i \xi_1^a \d_a \ph(x) ,
} and likewise for $J$ and $B_2$. One may express the $SO_0(2,1)$
generators in terms of Virasoro generators via \eqn{
  \label{eq:J}
  J   &=& L_0 - \Lt_0 , \\
  \label{eq:B1}
  B_1 &=& \half \left(L_1 + L_{-1} + \Lt_1 + \Lt_{-1} \right) , \\
  \label{eq:B2}
  B_2 &=& - \frac{i}{2} \left( L_1 - L_{-1} - \Lt_1 + \Lt_{-1} \right) .
}
We see that the de Sitter group is a diagonal subgroup of the
$SL(2,C)\times SL(2,C)$ generated by $L_0,\,L_{\pm 1},\,\Lt_0,\,\Lt_{\pm 1}$.

Constructing de Sitter-invariant states is non-trivial, as can be
seen from the expressions of the generators
(\ref{eq:J})-(\ref{eq:B2}). Because the boost generators contain
both raising and lowering Virasoro generators, it is difficult to
construct a non-trivial state that is boost invariant. Indeed, it is
easy to show that the only de Sitter-invariant state in our basis is
the $p=w=0$ vacuum $\ovac$ (see also \cite{Allen85}). Furthermore,
it can be shown that there exist no linear combinations of our basis
states that are both de Sitter-invariant and normalizable
\cite{Higuchi91a}. Thus $\ovac$ is the only de Sitter-invariant
state in the auxiliary Hilbert space $\cHa$.

\section{Group Averaging and the physical Hilbert space}
\label{sec:groupAveraging}

We now construct de Sitter invariant states via group averaging.  We
study the resulting physical Hilbert space $\cHp$ and provide an
orthonormal basis. We follow closely in the steps of
\cite{Higuchi91b} and,  in particular, define the space ${\cal
H}_{\rm seed} = \{\sket{\psi} \}$ of ``Higuchi seed states''
 which are:
\begin{enumerate}
\item[i)]
  $SO(2)$-invariant, i.e.
  \eq{\label{Jreq}
    J \sket{\psi} = 0 ,
  }
\item[ii)]
  annihilated by the lowering operators $L_1$ and $\Lt_1$,
  \eq{\label{Lreq}
    L_1 \sket{\psi} = \Lt_1 \sket{\psi} = 0 ,
  }
\item[iii)]
  in the subspace corresponding to eigenvalues $E > 1$ of $H$ (recall
  \ref{eq:H}).  We note that $H$ preserves the conditions
  \eqref{Jreq}, \eqref{Lreq} and so can be diagonalized in ${\cal
H}_{seed}$.
  \end{enumerate}
Furthermore, we will confine attention to a basis of such states
which are eigenstates of $E$ with inner products
  \eq{ \label{eq:normalization}
    \sbrak{\psi_1}{\psi_2}
    = \frac{(E-1)}{2} \delta_{\psi_1,\psi_2}.
  }
Here $\delta_{\psi_1,\psi_2}$ denotes the complete set of Kronecker
deltas needed to specify that $\sket{\psi_1}$ and $\sket{\psi_2}$
represent the same state in our basis  .

The group averaging of such states is easy to control. Criterion
(ii.) has the effect that \eq{
  B_1 \sket{\psi} = ( L_{-1} + \Lt_{-1} ) \sket{\psi},
  \quad
  \sbra{\psi} B_1 = \sbra{\psi} ( L_1 + \Lt_1 ) .
}
As a result $\sbra{\psi_1} B_1 \sket{\psi_2} = 0$ for all seed
states. Recalling the commutation relations
\eq{ \label{eq:VirasoroCommutators}
  \com{L_{\pm 1}}{L_0} = \pm L_{\pm 1} , \quad
  \com{L_1}{L_{-1}} = 2 L_0 ,
} we may compute $\sbra{\psi_1} (B_1)^2 \sket{\psi_2}$ by commuting
creation operators to the left (annihilation operators to the
right), with the result $\sbra{\psi_1} (B_1)^2 \sket{\psi_2} =
\sbra{\psi_1} (L_0 + \Lt_0) \sket{\psi_1} = E\,
\sbrak{\psi_1}{\psi_2}$, where $E$ is the energy of either state.
Continuing in this manner, one can readily see that \eqn{
  \sbra{\psi_1} (B_1)^m \sket{\psi_2}
  &=& 0 \quad m{\rm\; odd ,} \\
  \sbra{\psi_1} (B_1)^m \sket{\psi_2}
  &=& f(m,E) \, \sbrak{\psi_1}{\psi_2} \quad m{\rm\; even} ,
}
where $f(m,E)$ is a function of $m$ and $E$.
In particular, \cite{Higuchi91b} showed that
\eq{ \label{eq:HiguchiF}
  \sbra{\psi_1} e^{i\lambda B_1} \sket{\psi_2}
  = \left( \cosh \frac{\lambda}{2} \right)^{-2 E}
  \,\sbrak{\psi_1}{\psi_2}
} if the zero-mode is ignored, i.e. for the Fock space over the $p =
w = 0$ vacuum.  However, since dependence on $p,w$ enters only
through $E$, we see that \eqref{eq:HiguchiF} holds in general.
Working in terms of the the energy flux $E$ turns out to make the
inclusion of several effects of the zero mode quite straightforward.

We wish to compute the group averaging inner product of two seed
states. We begin by specializing expression
(\ref{eq:GAInnerProduct}) to the case of $SO_0(2,1)$. Using the
Cartan decomposition of $SO_0(2,1)$ we can write any group element
as a product of two $SO(2)$ rotations and a boost
\cite{Bargmann47,Vilenkin91}: \eq{
  U(g) = e^{i \alpha J} e^{i \lambda B_1} e^{i \gamma J} .
} Here $e^{i \alpha J}$ is the $SO(2)$ rotation through angle
$\alpha$ ($0 \le \alpha \le 2\pi$) and $e^{i\lambda B_1}$ is the 
boost along $\xi_1^a$ with rapidity $\lambda$ ($0 \le \lambda \le \infty$). 
In a similar fashion, the Haar measure can be decomposed as 
\eq{
   dg = \frac{1}{4\pi^2} \,d\alpha\, d\gamma\, d\lambda\, \sinh \lambda ,
}
where $\frac{1}{2\pi} d\alpha$ and $\frac{1}{2\pi}d\gamma$ are both
Haar measures on $SO(2)$. The group averaged inner product is then
\eqn{
  \label{eq:innerProduct1}
  \brak{\Psi_1}{\Psi_2} &=& \frac{1}{4\pi^2}
  \int d\alpha\, d\gamma\, d\lambda \, \sinh \lambda \,
  \dbra{\psi_1} e^{i \alpha J} e^{i \lambda B_1} e^{i \gamma J}
  \dket{\psi_2} \nn \\
  &=&
  \int_0^\infty d\lambda \, \sinh \lambda \,
  \dbra{\psi_1} \Proj \, e^{i\lambda B_1} \Proj \dket{\psi_2} .
}
In the second line we have identified the projector $\Proj$ onto
$SO(2)$-invariant states
\eq{
  \Proj = \frac{1}{2\pi} \int_0^{2\pi} d\alpha \, e^{i\alpha J} .
}

Now consider the inner product of two physical states built
from seed states $\ket{\Psi_{1,2}} := \int dg U(g) \sket{\psi_{1,2}}$.
From (\ref{eq:innerProduct1}) we have
\eqn{
  \label{eq:innerProduct2}
  \brak{\Psi_1}{\Psi_2}
  &=&
  \int_0^\infty d\lambda \, \sinh \lambda \,
  \sbra{\psi_1} \Proj \, e^{i\lambda B_1} \Proj \sket{\psi_2} \nn \\
  &=&
  \int_0^\infty d\lambda \, \sinh \lambda \,
  \sbra{\psi_1} e^{i\lambda B_1} \sket{\psi_2} \nn \\
  &=&
  \sbrak{\psi_1}{\psi_2}
  \int_0^\infty d\lambda \, \sinh \lambda
  \left( \cosh \frac{\lambda}{2} \right)^{-2 E} \nn \\
  &=&
  \sbrak{\psi_1}{\psi_2} \, \frac{2}{(E-1)} \nn \\
  &=& \delta_{\psi_1,\psi_2} .
} In the second line we used the fact that seed states are
$SO(2)$-invariant; in the third line we used (\ref{eq:HiguchiF}).
Evaluating the integral and inserting our normalization
(\ref{eq:normalization}) leads to the final result. This is
essentially the same calculation as was performed previously in
\cite{Higuchi91b} without the zero mode. We emphasize again that our
formalism allows a quick generalization to the case of non-vanishing
zero-mode.

We see from (\ref{eq:innerProduct2}) that the set of de
Sitter-invariant states $\{\ket{\Psi_i}\}$ built from the seed
states $\{ \dket{\psi_i}\}$ forms an orthonormal set. One can also
show that this set spans the space of de Sitter-invariant states
constructed from linear combinations of auxiliary states with $E >
1$. The proof is exactly as in \cite{Higuchi91b}.  Thus we have an
orthonormal basis of states as desired.

We conclude this section with a discussion of auxiliary states with
$E \le 1$. It is natural to ask whether such states contribute to
the physical Hilbert space and, if so, how we can incorporate them
into our formalism. Fortunately, for any vacuum there are only a few
such states (and there are none for $h + \tilde h > 1$). In
particular,
\begin{enumerate}

  \item[]{\bf The case $p=w=0$:}
    The only states with $E \le 1$ are the vacuum $\ovac$ and the
    single-particle states $\a_{-1}\ovac$, $\at_{-1}\ovac$. The
    vacuum is de Sitter-invariant and can be a state in the
    physical Hilbert space. However, this state must be treated separately since for $\ovac$ group
    averaging does not converge.  This separate treatment may be
    justified via the observation (see \cite{ALMMT,Single}) that $\ovac$
     is superselected from states where group averaging does converge.  Turning now to
     the single-particle states, one notes that they each have angular momentum $\pm1$.
     As a result, the group average of such states over the rotation group SO(2) $\subset$ SO(2,1)
     already vanishes and we do not expect these states to contribute the physical Hilbert space.

  \item[]{\bf The case $p^2 \le 1, w = 0$,
      or $p=0, R^2 w^2/4 \le 1$:} Here the only states with $E \le 1$ are
    vacua $\hvac$.  Such vacua are \emph{not} de Sitter invariant,
    though group averaging again fails to converge.   In this case we
    expect an appropriately renormalized form of group averaging to
    converge, though we leave this for future work.  The resulting de
    Sitter-invariant states will again be superselected from states for
    which no such renormalization was needed.   See
    \cite{ALMMT,Single,GomberoffMarolf98,JLAM03,JLAM05} for
    further examples and discussion of this phenomenon.
    
  \item[]{\bf The case
      $|k| = |w| = 1$ and $R = \sqrt{2}$:}  Such states also have $J = \pm
    1 \neq 0$ and again group average to zero under SO(2) $\subset$
    SO(2,1).  We expect no physical states from such seed states.
    
\end{enumerate}

\section{Physical Entropy}
\label{sec:entropy}

We now compute the density of physical states.  One typically
computes this density as a function of energy.  However, as
previously remarked, there is no natural conserved notion of energy
in de Sitter space.  Moreover, those charges which are associated
with de Sitter isometries must vanish for physical states.  We will
thus need to find some other notion of energy to use below.

A natural approach is to follow \cite{GiddingsMarolf07} and to
consider the energy $E$ defined in section \ref{Sec:merge}, which
measures the flux of stress energy through the surface $\tau =0$.
Since the definition of $E$ as the eigenvalue of $H$ is not de
Sitter invariant, this quantity is not a priori defined on physical
states. Nevertheless, a de Sitter-invariant notion of $E$ was
defined in \cite{GiddingsMarolf07}. In our present language, the de
Sitter invariant energy is the operator whose eigenstates are
precisely the physical states $| \Psi_i \rangle$ obtained by group
averaging Higuchi seed states $\sket{\psi_i}$, and such that the
eigenvalue of $| \Psi_i \rangle$ is just the eigenvalue of $H$ for
$\sket{\psi_i}$. The one to one map between physical states and seed
states, and that fact that the $| \Psi_i \rangle$ form an orthogonal
basis of $\cHp$, imply that this de Sitter-invariant energy is a
self-adjoint operator on $\cHp$.  Furthermore, because any two states
related by the action of some $U(g)$ yield the same state under
group averaging, we see that defining $H$ in some other reference
frame (i.e., replacing $H$ by $U(g) H U(g^{-1})$for some $g$ in the
definition of a Higuchi seed state) would lead to the same de
Sitter-invariant notion of energy.  In a very rough sense, this
energy operator considers the energy flux of a physical state
through each possible de Sitter neck (associated with each possible
choice of reference frame) and reports the smallest value obtained.
For simplicity, we will again use $E$ to denote the eigenvalue of
this de Sitter-invariant energy.

It is clear that counting the density of physical states is
equivalent to counting the number of Higuchi seed states as a
function of $E$.  The density diverges when the scalar target space
is non-compact, so we focus on the case with $S^1$ target space.  As
usual, we perform the calculation separately for the Fock space over
each vacuum $\hvac$. Our task is thus to compute $\ln \cNs(N)$, the 
logarithm of the number of Higuchi seed states as a function of the 
level $N$ (recall $N = E - h - \h$) above each vacuum $\hvac$.
To do so, we must first examine the seed state criteria in more
detail. We begin with $SO(2)$ invariance which requires \eq{
  J \hnket = (kw + n + \nt ) \hnket = 0 ,
 }
 so that  an $SO(2)$-invariant state at level $N$ must satisfy \eq{
\label{eq:n}
  n = \half( N - kw ), \quad \nt = \half (N + kw) .
}
 Since $n$ and $\nt$ must be non-negative integers,
$SO(2)$-invariant states are possible only at levels $N = k w + 2
m$, where $m$ is a non-negative integer.

Next consider the property $L_1 \sket{\psi} = 0$. Because the left-
and right-moving sectors commute, we can decompose the Fock space
over a given vacuum into left- and right-moving Hilbert spaces $\cHa
= \cHl \otimes \cHr$; an auxiliary state with levels $(n,\nt)$ is
then in the product space $\cH_{n} \otimes \cH_{\nt}$. Because $L_1$
acts only on left-movers we can focus on $\cH_n$. The annihilation
operator $L_1$ lowers $n$ by $1$, i.e. $L_1 \hnket =
\dket{n-1,\nt;h,\h}$. In fact, $L_1$ is a surjective map from
$\cH_n$ to $\cH_{n-1}$: \eq{
  L_1 : \cH_{n} \to \cH_{n-1} .
}
The number of states in $\cH_n$ annihilated by $L_1$ is therefore given
by the difference in dimension
\eqn{
  \label{eq:counting}
  \left(
  \begin{array}{c}
    {\rm \#\;states\;in\;}\cH_n \\
    {\rm annihilated\;by\;} L_1
  \end{array}
  \right)
  &=& \dim( \cH_n ) - \dim( \cH_{n-1} ) = P(n) - P(n-1) ,
} where $P(x)$ is the number of integer partitions in $x$
\cite{Ginsparg88}. For $n=1$ we have $P(1) - P(0) = 0$, while for
$n>1$ we have $P(n) - P(n-1) > 0$. The same argument applies for the
action of $\Lt_1$ on the right-moving sector $\cH_{\nt}$. Combining
our observations (\ref{eq:n}) and (\ref{eq:counting}), we find that
the number of seed states at level $N$ is \eqn{
  \label{eq:numberStates}
  \cNs(N)
  &=& P\left(\frac{N+kw}{2}\right)
  \left[P\left(\frac{N-kw}{2}\right)
    - P\left(\frac{N-kw-2}{2}\right)\right] \nn \\ & &
  \times P\left(\frac{N-kw}{2}\right)
  \left[P\left(\frac{N+kw}{2}\right)
    - P\left(\frac{N+kw-2}{2}\right)\right].
}
One can easily re-write this as a function of the energy flux $E$, though in practice this will not be required to establish agreement with the density of states in $\cHa$.

Let us now compare the number of seed and auxiliary states for
both small and large $N$. Note that the number of states in $\cHa$ with given $(n,\nt, h, \tilde h)$ is
simply $\cNa(n,\nt) = P(n)P(\nt)$ and the total number of states
at level $N$ is
\eq{\label{Naux}
  \cNa(N) = \sum_{n' = 0}^N P(n') P(N-n') .
} Because $P(x) \sim x$ for $x$ of order 1, for small $N$ there are
dramatically fewer seed states than auxiliary states: \eq{
  \cNs(N \sim 1) \sim N , \quad
  \cNa(N \sim 1) \sim N^2 .
}
However, it is more interesting to compare entropies in the
thermodynamic limit of large $N$.
For $N \to \infty$ we may use the Hardy-Ramanujan formula for the
asymptotic behavior of $P(x)$\cite{Hardy}:
\eq{
  \label{eq:Hardy}
  P(x) \approx \frac{1}{4 x \sqrt{3}}
  \exp\left[ 2 \pi \sqrt{\frac{x}{6}} \right]
  \;\;{\rm as}\; x \to \infty .
}
Inserting \eqref{eq:Hardy} into \eqref{eq:numberStates} and using Cardy's
formula \cite{Cardy86} to compute \eqref{Naux}
yields
\eq{
  \Ss(N) \approx \Sa(N) \approx \frac{2 \pi}{\sqrt{3}} \sqrt{N},
}
so that  in this limit the seed state entropy agrees with the entropy of
the auxiliary Hilbert space as claimed.  This provides an explicit confirmation of the general argument
given in the appendix of \cite{GiddingsMarolf07}.

\section{Discussion}
\label{sec:discussion}

We have studied the behavior of 1+1 massless scalars under de Sitter
group averaging, building on earlier work by Higuchi
\cite{Higuchi91b}.  The new element was to include the scalar zero
mode.   We constructed an orthonormal set of de Sitter-invariant
states which forms a basis of the physical state space (up to the
minor exceptions discussed in section \ref{sec:groupAveraging}).  We
have also computed the entropy of this physical space.   As
anticipated in \cite{GiddingsMarolf07}, to leading order at large
$E$ this entropy agrees with the entropy of the auxiliary Hilbert
space.  This observation supports the claim that group averaging
will yield enough states to reproduce classical physics in the
$\hbar \rightarrow 0$ limit.

The elegant orthonormal bases constructed constructed here and in 
\cite{Higuchi91b} belie the fact that implementing group averaging for 
more general fields can be quite difficult
 (see e.g.\cite{MarolfMorrison08}). The simplicity
of group averaging for the present model is in part due working in
low dimensions, but has more to do with the presence of conformal
symmetry. For general masses and dimensions one looses the useful
commutation relations between the raising and lowering parts of
$B_1$ (here $L_1 + \tilde L_{1}$ and $L_{-1} + \tilde L_{-1}$) and
$E$ which in our case followed from the Virasoro algebra. Similar
relations do hold, however, for conformally-coupled scalar fields in
arbitrary dimension, i.e. for scalars on $dS_{d+1}$ which satisfy 
the equation of motion
\eq{
  g^{ab}\nabla_a\nabla_b \ph(x) = \left(\frac{d^2-1}{4}\right)\ph(x). 
}
In such cases it is straightforward to use Higuchi's algorithm
to construct an orthonormal basis of physical states. (The 3+1 case
was studied in \cite{AHnotes}.) Additionally, in the right choice of
gauge, both free gravitons and free gauge vector fields in $3+1$
dimensions have boost matrix elements identical to those of
conformally coupled scalar fields, and one may again construct an
orthonormal basis \cite{Higuchi91b}. 

The more general case remains open for future work. However, one 
expects the massless scalar field in higher dimensions to be 
qualitatively similar to the case discussed here. I.e., we expect that 
there is some analogue of our quantity $E$ which in some sense measures 
the total excitation of the state, including contributions from both 
particles and the zero mode.  One expects group averaging to fail 
when the quantity is very small, but to converge when it is 
sufficiently large.

\subsection*{Acknowledgements}

The authors are grateful to Atsuchi Higuchi for many discussions of
group averaging and for sharing his unpublished notes
\cite{AHnotes}. This work was supported in part by the US National
Science Foundation under Grant No.~PHY05-55669, and by funds from
the University of California.


\end{document}